\documentclass[11pt, onecolumn]{article}
\usepackage[paper=a4paper, hscale=0.8, vscale=.8]{geometry}
\usepackage{graphicx}
\usepackage{amssymb}
\usepackage{amsmath}
\usepackage{bbm}
\usepackage{url}
\usepackage{simplewick}
\usepackage{prettyref}
	\newcommand{\pr}[1]{\prettyref{#1}}
	\newrefformat{app}{appendix~\ref{#1}}
	\newrefformat{eqn}{Eq.\,(\ref{#1})}
	\newrefformat{eqns}{Eqs.\,(\ref{#1})}
	\newrefformat{sec}{section~\ref{#1}}
	\newrefformat{subsec}{section~\ref{#1}}
	\newrefformat{subsubsec}{section~\ref{#1}}
	\newrefformat{fig}{fig.~\ref{#1}}
\usepackage[usenames,dvipsnames]{color}
\usepackage{cite}
\newcommand{\V}[1]{\mathbf {#1}}

\newcommand{\todo}[1]{}
\newcommand{\chdid}[1]{}

\newcommand{\notes}[1]{}
\newcommand{\js}[1]{}
\newcommand{\SimplePhoton}[2]{
\raisebox{#1}{
	\includegraphics[width=#2]{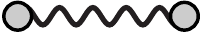}
}
}
\newcommand{\DressedPhotonE}[2]{
\raisebox{#1}{
	\includegraphics[width=#2]{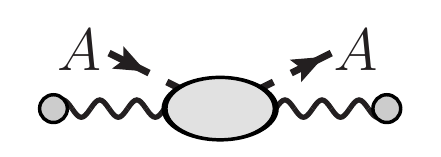}
}
}
\newcommand{\DressedPhotonG}[2]{
\raisebox{#1}{
	\includegraphics[width=#2]{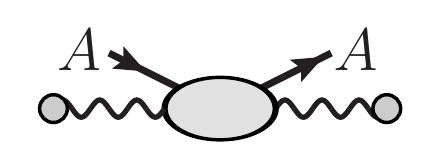}
}
}
\newcommand{\GPropOneLoop}[2]{
\raisebox{#1}{
	\includegraphics[width=#2]{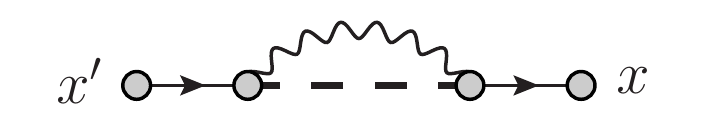}
}
}
\newcommand{\SigmaG}[2]{
\raisebox{#1}{
	\includegraphics[width=#2]{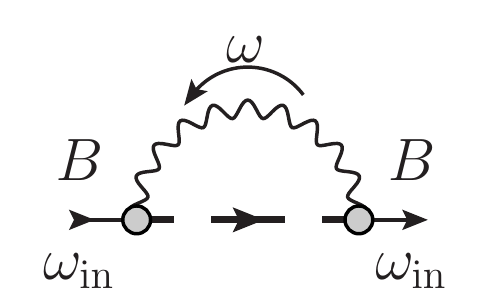}
}
}
\newcommand{\BoxStandard}[2]{
\raisebox{#1}{
	\includegraphics[width=#2]{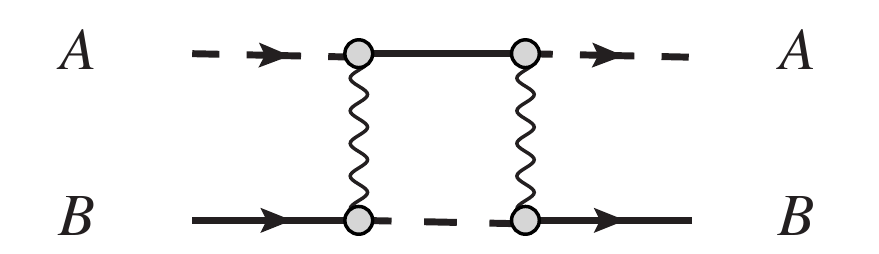}
}
}
\newcommand{\CrossStandard}[2]{
\raisebox{#1}{
	\includegraphics[width=#2]{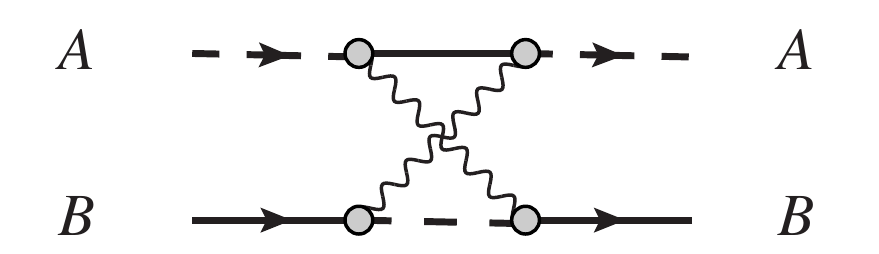}
}
}
\newcommand{\BoxRightEar}[2]{
\raisebox{#1}{
	\includegraphics[width=#2]{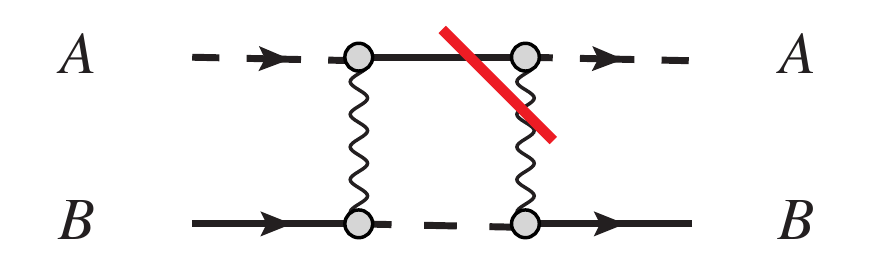}
}
}
\newcommand{\CrossRightEar}[2]{
\raisebox{#1}{
	\includegraphics[width=#2]{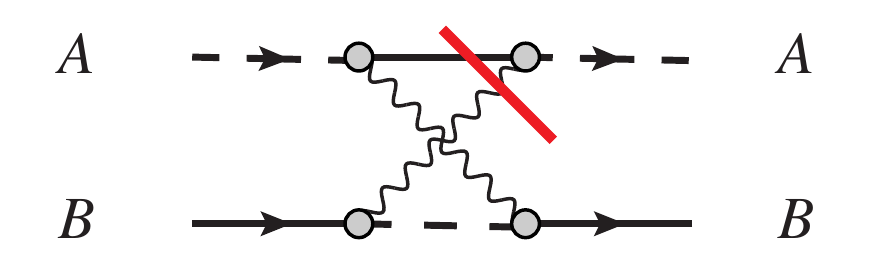}
}
}
\newcommand{\ResonanceInteraction}[2]{
\raisebox{#1}{
	\includegraphics[width=#2]{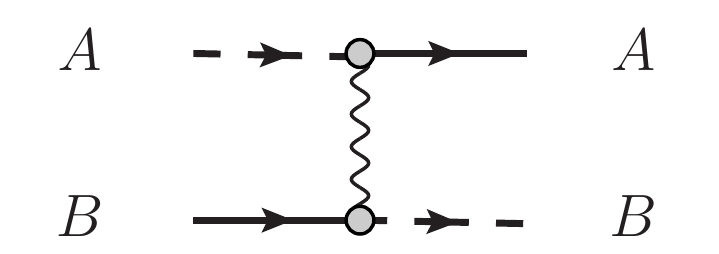}
}
}
\newcommand{\BoxGG}[2]{
\raisebox{#1}{
	\includegraphics[width=#2]{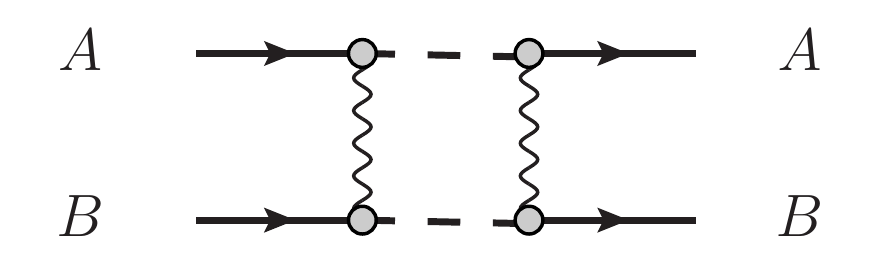}
}
}
\newcommand{\CrossGG}[2]{
\raisebox{#1}{
	\includegraphics[width=#2]{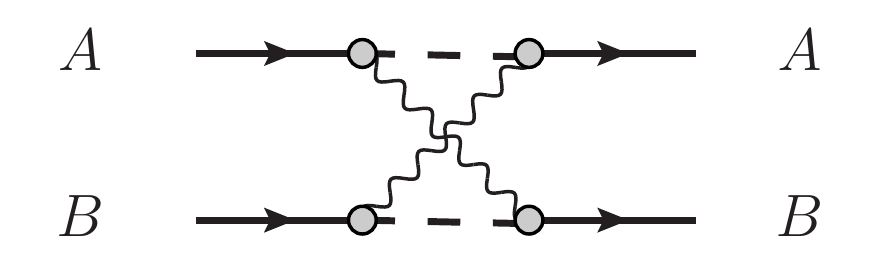}
}
}

\newcommand{\refcite}[1]{\cite{#1}}

\begin{document}


\title{Feynman Diagrams for Dispersion Interactions Out of Equilibrium --\\
Two-Body Potentials for Atoms with Initial Excitation} 

\author{
\normalsize Harald R. Haakh$^1$, J\"urgen Schiefele$^{1, 2}$, Carsten Henkel$^1$
\bigskip
\\
\normalsize $^1$Institut f\"ur Physik und Astronomie, Universit\"at Potsdam,\\
\normalsize  Karl-Liebknecht-Str. 24/25, 14476 Potsdam, Germany\\
%
%
\normalsize $^2$Departamento de F\'isica de Materiales,\\
\normalsize  Universidad Complutense de Madrid,
\normalsize  28\,040 Madrid, Spain}

\maketitle


\begin{abstract}\noindent
Diagrammatic techniques are well-known in the calculation of dispersion interactions between atoms or molecules. The multipolar coupling scheme combined with Feynman ordered diagrams significantly reduces the number of graphs compared to elementary stationary perturbation theory. We review calculations of van der Waals-Casimir-Polder forces, focusing on two atoms or molecules one of which is excited. In this case, calculations of the corresponding force are notorious for mathematical issues connected to the spontaneous decay of the excitation.
Treating such unstable states in a full non-equilibrium theory provides
a physical interpretation of apparent contradictions in previous results and underlines the importance of decay processes for the intermolecular potential.
This may have important implications on reactions in biological systems, where excited states may be relatively long-lived and the resonant intermolecular force may result in directed Brownian motion. 

\end{abstract}

\section{Tools from Non-equilibrium Field Theory in Molecular QED}
The interaction between electrically neutral yet polarizable particles, 
commonly named after J. D. van der Waals, F. London, H. Casimir, and D. Polder,
is one of the fundamental problems in atomic and molecular physics.
While the two-body potential between two particles (atoms, molecules, nanoparticles) 
in their internal ground state 
is unambiguous\cite{Casimir_1948,Power_1983,McLachlan_1963,Craig_1998}, apparently incompatible results have been obtained 
if one of the atoms is prepared in an excited state. 
The quantity of interest is the long-range part of the potential 
that overwhelms the familiar van der Waals interaction:
some calculations found this part to oscillate spatially%
\cite{McLone_1965,Philpott_1966,Kweon_1993},
while later work found a monotonic power law%
\cite{Power_1993}\nocite{Power_1995,Sherkunov_2007,Sherkunov_2009}.
It is the aim of the present paper to understand these differences\cite{Power_1995,Sherkunov_2007} better.
We handle this non-equilibrium problem with a diagrammatic (Feynman graph) expansion.

The starting point of our quantum mechanical description is the electric dipole coupling Hamiltonian
\begin{align}
H_{AF} 
	= 
	-
	\sum_{n} E_i(x_n) 
	\left[d_i^{n}\psi^e(x_n) \psi^{g\dagger}(x_n) 
	+ d_i^{n *}\psi^g(x_n) \psi^{e\dagger}(x_n)\right]
~.
	\label{eq:dipole-coupling-Hamiltonian}
\end{align}
%
For simplicity, we treat the atoms as pointlike two-level objects located 
at the spacetime coordinates $x_n = \{\V{r}_n, t_n\}$.
The fermionic operators $\psi^{a\dagger}(x)$ create atoms in state
$a= g, e$ at $x$. Assuming that the transition dipoles $d_i^n$ are
real, we work in the following with an effectively scalar electric field
$E( x ) = d_i^n E_i ( x )$ whose space argument accounts for possible
differences in the transition dipoles.

The textbook treatment of dispersive atom-atom 
interactions\cite{Craig_1998,Salam_2009}
employs stationary fourth-order perturbation theory in $H_{AF}$.
The summation over all possible intermediate states leads to a rather large 
number of terms.
In this work, we discuss a more concise treatment
using the closed time-path contour formalism due to Schwinger, Craig, Mills, 
and Keldysh (see Ref.\,\refcite{Danielewicz_1984}).
Following Sherkunov 
\cite{Sherkunov_2007,Sherkunov_2009,Sherkunov_2005},
the level shift of a ground-state atom is extracted from its full propagator
\chdid{[$T$ order symbol removed]} \todo{HH: \Large$\checkmark$}
\begin{align}
i  g^{\rm (full)}_{\alpha \beta}(x,y)
	&\equiv	
	\langle
	T_c \bigl\{
	S_2(-\infty, \infty) 
	S _1(\infty, -\infty)
	\psi^g_\alpha (x) \psi^{g \dagger}_\beta (y)
	\bigr\}
	\rangle^{\rm conn}
	~.
	\label{eqn:Dyson_Feynman}
\end{align}
The greek subscripts take the values 1 or 2,
denoting the branch of the Keldysh time contour
which extends from $t=-\infty$ to $\infty$ (branch 1) and back (branch 2).
The prescription $T_c\{ \dots
\}$ orders operators according to their time arguments on the contour.
Finally, 
$S$ is the standard scattering operator,
the brackets $\langle \dots \rangle$ denote 
an eigenstate of the non-interacting theory%
\chdid{[CH: need a more general state here, since the excited atom A will
come in]} \todo{HH: \Large $\checkmark$}%
,
and only \emph{connected} products contribute.
The Feynman formalism\cite{Schiefele_2010,Schiefele_2011} is valid for known in- and out-states, 
which is the typical scenario 
in equilibrium systems at zero temperature.
If, however, a general initial state 
(including, in our case, unstable excited atomic states) 
is prepared and then left to evolve under the interaction, 
the Keldysh formalism is necessary to describe all energetically allowed processes.

In the following, a propagator with both times on the forward 
branch will be used. Its series expansion is from
Eq.\,(\ref{eqn:Dyson_Feynman})
\chdid{[prefactor $-1$ added! not $c$-integration!]}\todo{HH:\Large $\checkmark$}
\begin{align}
& i  g^{\rm (full)}_{1 1}(x,y)
	=
	\langle
	T \bigl\{
	\psi^g (x) \psi^{g \dagger} (y)
	\bigr\}
	\rangle
\nonumber
\\
	&-
	\frac{1}{2}
	\int d{t_1} d{t_2} \,
	\langle
	T_c \bigl\{
	\sum_{\alpha,\beta = 1}^2\, 
	(-1)^{\alpha + \beta}
	H_{AF, \alpha} (t_1) H_{AF, \beta}(t_2) 
	\psi^g_1 (x) \psi^{g \dagger}_1 (y)
	\bigr\}
	\rangle^{\rm conn}
	+ 
	\dots
\;.
	\label{eq:Dyson-Keldysh-11}
\end{align}
where the first line is the ordinary (bare) Feynman propagator given
in Eq.\,(\ref{eq:g-propagator}), with $T$ being the standard time ordering.
We will see below that processes involving only ground-state atoms can be entirely
described in terms of Feynman propagators.


The next section presents the calculation of self-energies 
for atoms and photons, and recovers the energy shifts and decay widths in a 
system of two particles. A comparison of the Feynman and Keldysh results
suggests an interpretation of the apparent disagreement of earlier
results: these may apply in different time regimes after preparation, separated 
by the time scale that characterizes energy transfer by an exchange of
resonant photons.
Finally, we discuss how resonant long-range potentials might influence some systems of biological relevance that are not commonly considered in the community present at this conference.
%
%
%
%
%
\section{Self-Energies with Dressed Photons}
Our perturbative analysis is built from two basic expressions:
the self-energies of an atom coupled to the photon field 
and of  a photon coupled to a (second) atom, respectively.
Conceptually, this is closely related to the notion of dressed states\cite{Compagno_1995}.
The real and imaginary parts of the self-energy 
\chdid{[CH: minus sign in decay rate]}
\begin{align}
\Sigma_{1 1} = \Delta E - i \frac{ \Gamma }{2} 
\label{eqn:sigma_re_im}
\end{align}
can be identified as the energy-shift and the inverse lifetime (decay width) of the particle.
We first calculate the atomic self-energy at next-to-leading order in the
coupling to a generic photon field. 
In a second step, the interatomic interaction is identified by ``dressing'' the photons with a second atom.
We only have to consider terms where photons are connecting the two atoms,
provided we assume the transition frequencies to include the single-particle 
Lamb-shift.
%
%
%
%
%
\subsection{Single atom plus field: atom self-energy}
\label{sec:atom-is-dressed}
%
%
%
We start by considering the propagator of a ground-state atom $B$ 
(transition frequency $\omega^B$)
at the one-loop level. 
Using the Feynman rules from the Appendices and App. C of 
Ref.\,\refcite{Schiefele_2011}, the evaluation of
Eq.\,(\ref{eq:Dyson-Keldysh-11}) yields
\chdid{[CH: factor $i$ removed, from expansion of (\ref{eq:g-and-e-propagator-11})
around $\epsilon_g$;
consistent with Eq.\,(22) of Sherkunov2007]}
\begin{align}
g_{1 1}^{(2)} &(x,x')
	=
	\GPropOneLoop{-0.ex}{0.25\textwidth}
\nonumber
\\
	&=
	\int d^3 x_1
	\int d^3 x_2
	\int d \omega
	\, e^{-i\omega(t - t')}
	g_{1 1} (\omega, \V{r},\V{r}_1)
	\Sigma_{11}^{g} (\omega, \V{r}_1,\V{r}_2)
	g_{1 1} (\omega, \V{r}_2,\V{r}')
\;,
\label{eqn:def_selfenergy}
\end{align}
%
where the self energy for a pointlike atom is to lowest order
\begin{align}
\Sigma_{1 1}^{g B} ( \omega_{\rm in} )
	&= 
	\SigmaG{-2ex}{0.17\columnwidth}
	= 
	-i
	\int 
	\frac{d\omega}{2\pi} \, 
	D_{1 1}(\omega, \V r_B, \V r_B)
	e_{1 1}( \omega_{\rm in} + \omega )
\;.
	\label{eqn:Sigmag}
\end{align}
The excited state propagator $e_{1 1}$ is given in 
Eq.\,(\ref{eq:g-and-e-propagator-11}) and $D_{1 1}$ denotes the Feynman 
propagator for photons. For the self-energy
$\Sigma^e_{1 1}$ of an excited atom, replace $e_{1 1}$ by
$g_{1 1}$ in the integral.
%
This result can be brought into a more familiar form (see 
Ref.\,\refcite{Wylie_1985}) by evaluating it
on the mass shell, $\omega_{\rm in} = \epsilon^B_g$, and writing the
integral over positive frequencies only (recall that 
$D_{1 1}(\omega)$ is even in $\omega$):
\begin{align}
\Sigma^{g B}_{1 1}( \epsilon^B_g )
	&= 
	i
	\int_0^\infty 
	\frac{d\omega}{2\pi} \, 
	D_{1 1}(\omega, \V r_B, \V r_B) 
	\alpha_{1 1}^{g B} (\omega)
\label{eqn:sigmaGp} 
\;.
\end{align}
Here we recover the Feynman-ordered polarizability
%
\begin{align}
\alpha_{1 1}^{g B} (\omega)
	&= 
	\frac 1 {\omega^{B} - \omega - i\, 0^+}
	+
	\frac 1 {\omega^{B} + \omega - i\, 0^+}
	~,
\label{eqn:alpha11} 
\end{align}
that is evaluated explicitly in \ref{sec:polarizations}.
%
%
%
(For the excited two-level atom $A^*$, replace $\omega^B \to -\omega^A$.)
Note that the propagators here
have poles in the upper left and lower right quadrant of the complex
frequency plane 
(Feynman prescription). They coincide, however, with the usual retarded 
response functions at positive frequencies, and only these appear in
the complex self-energy~(\ref{eqn:sigmaGp}).

%
%
Substituting the bare photon propagator
into $\Sigma^{g/e}_{1 1}$
yields, according to \pr{eqn:sigma_re_im}, an infinite energy shift which is, 
of course, the unrenormalized Lamb shift of atom $B$.
The imaginary part of $\Sigma^{g}_{1 1}$ is nonzero if the temperature 
$T > 0$ [see Eq.\,(\ref{eq:photon-11-finite-T})], describing the absorption rate
of thermal photons. At $T = 0$, an imaginary part is found for the excited
atom $A^*$: this comes from
a pole of $\alpha^{e A}_{1 1}( \omega )$ in the upper right quadrant and
gives the spontaneous decay rate
%
$\Gamma^{e A}_0 = 2 \,{\rm Im}\, D_{1 1}(\omega_A, \V r_A, \V r_A ).$
%
%
%
%
\subsection{Photon propagation in the presence of a second atom}
\label{sec:NLO_photon}
To evaluate the impact of a second atom (labeled $A$) on the self-energy of
atom $B$, we take the diagram~(\ref{eqn:Sigmag}) and replace the
photon propagator by its next-to-leading order correction:
$$
\SimplePhoton{-0.1ex}{0.08\textwidth}
\;\to\;
\DressedPhotonG{-1.ex}{0.15\textwidth}
{\rm or}
\DressedPhotonE{-1.ex}{0.15\textwidth}
~,
$$
depending on the state of atom $A$. This ``modular'' strategy is, of
course, consistent with the fourth-order expansion of the 
atom propagator $g_{1 1}$ in Eq.\,(\ref{eqn:Dyson_Feynman}).
\todo{[CH: should be true, but did somebody check? HH: That is how we compared to the ear-diagrams, so: \Large $\checkmark$]}
%
The bare photon propagator $D_{1 1}( x, x' )$ is given in 
\ref{sec:free_photons}, and its correction to leading order 
recovers Eq.~(28) of Ref.\,\refcite{Sherkunov_2007} 
\todo{HH: Do we really need this reference?}
\begin{align}
D^{(2)a}_{1 1} (x', x)
	&=
	\frac{- i}{2}
	\int\! d t_1 d t_2 \,
	\langle T_c \bigl\{
	\sum_{l, l' = 1}^2
	(-1)^{l+l'}
	H_{AF,l}(t_1) H_{AF,l'}(t_2)
	E_1(x') E_1(x)
	\bigr\}
	\rangle_{ a }^{\rm conn}~. 
\nonumber
\end{align}
%
where $a = e, g$ denotes the state of atom $A$.
For the purpose of our analysis it is sufficient to restrict the field to zero temperature
which simplifies the treatment in the frequency domain considerably
(see~\ref{sec:free_photons}).
We find from Wick's theorem\cite{Danielewicz_1984} 
\todo{[CH: please check space arguments and minus sign. HH: Looks good \Large$\checkmark$]}
\begin{eqnarray}
D^{ (2) a }_{1 1}(\omega > 0, \V r', \V r) 
		&= & D_{1 1}(\omega, \V r', \V r_A ) \alpha^{ a A }_{1 1}(\omega)
		D_{1 1}(\omega, \V r_A, \V r )  
		\nonumber\\
		&& {} 
		- D_{1 1}(\omega, \V r', \V r_A) \alpha^{ a A }_{1 2}(\omega)
		D_{2 1}(\omega, \V r_A, \V r) 
	\label{eq:D11-dressed-photon}
\end{eqnarray}
where the atomic polarizabilities are given in \pr{sec:polarizations}.
Note that this contains an extra term that is not in the form of Feynman
propagators. 
There is no connected diagram involving $D_{2 2}$, and 
at positive frequencies, $D_{1 2}(\omega) = 0$ 
(see~\ref{sec:free_photons}).
%
%
%
%
%
\subsection{Two ground-state atoms}
If atom $A$ is in the ground state ($a = g$), the contribution from the 
second Keldysh branch vanishes because $\alpha^{g A}_{1 2}( \omega > 0 ) 
= 0$.
We obtain a two-atom self-energy by substituting the dressed photon 
propagator $D^{(2)g}_{1 1}$ into $\Sigma^{g}_{1 1}$,
\todo{[CH: factor $i$ added, please check. HH: Is this a global prefactor for all self-energies?]}
\begin{align}
\Sigma^{gg}_{1 1} 
	&=
	\BoxGG{-2.5ex}{0.25\columnwidth} 
	+ 
	\CrossGG{-2.5ex}{0.25\columnwidth}
\nonumber
\\
	&=
	i
	\int_0^\infty \frac{d \omega}{2 \pi }\alpha^{g B}_{11}(\omega)
	\alpha^{g A}_{1 1}(\omega)
	D_{1 1}(\omega, \V r_A, \V r_B)
	D_{1 1}(\omega, \V r_B, \V r_A)
	~.
\nonumber
\end{align}
Only Feynman-ordered quantities contribute, and indeed, this situation corresponds
to the ground state of the non-interacting theory where no non-equilibrium
formalism is needed.
The integration lends itself to a rotation to imaginary frequencies, and one
sees that $\Sigma^{g g}_{1 1} = \Delta E^{g g }$ is a purely real 
two-body potential, equal to the well-known Casimir-Polder potential if the
two atoms are embedded in free space.
Note that the same calculation in traditional perturbation theory, e.g. in 
Ref.\,\refcite{Craig_1998}, 
involves 12 diagrams with time-directed photon-lines rather than the two 
Feynman diagrams above. 

\subsection{Ground-state atom and excited atom}

The additional term in the photon propagator~(\ref{eq:D11-dressed-photon})
makes the Feynman and Keldysh results differ. We find the following
self-energy for the ground-state atom $B$ in the presence of an 
excited atom $A^*$
\js{Probably we don't need to put labels on the points here.}
\todo{[CH: factor $i$ added and position arguments corrected, please check.]}
\begin{align}
\Sigma^{ge}_{1 1} 
	&=& i \int_0^\infty \frac{d \omega}{2 \pi } 
	\alpha^{g B}_{11}(\omega)
	\left\{\right.
 D_{1 1}(\omega, \V r_B, \V r_A) 
 	\alpha^{e A}_{1 1}( \omega ) 
	D_{1 1}(\omega, \V r_A, \V r_B)  &&
 	\nonumber
	\\
&& \left. {}- D_{1 1}(\omega, \V r_B, \V r_A) 
	\alpha^{e A}_{1 2}( \omega ) 
	D_{2 1}(\omega, \V r_A, \V r_B) 
	\right\}&&
\label{eqn:dressed_photon_2} 
\\
\nonumber%
&=& \BoxStandard{-2.5ex}{0.25\columnwidth} + \CrossStandard{-2.5ex}{0.25\columnwidth}\nonumber&&\\
&& +\BoxRightEar{-2.5ex}{0.25\columnwidth} + \CrossRightEar{-2.5ex}{0.25\columnwidth}&&
\label{eqn:dressed_photon_2a} 
\end{align}
The first line is the result obtained in Refs.\,\refcite{McLone_1965}--\refcite{Kweon_1993}.
In our formalism, this part of the result contains only time-ordered (Feynman) quantities and is represented
by the first two diagrams of \pr{eqn:dressed_photon_2a}.
We will argue in the next section under which conditions this line gives already the full result.

The second set of diagrams illustrates the terms that arise from
the second Keldysh branch. There is one vertex beyond the (thick red) bars
that corresponds to an interaction operator $H_{AF,2}( x )$ 
on the second Keldysh branch. It will end leftmost after the contour ordering
$T_c$ and can be interpreted as acting directly on the outgoing states. 
This modification of the out-state translates the equilibration of the 
initial state (prepared at $t = - \infty$ and left to evolve under the interaction). 
The thick red bar thus illustrates
the split between the branches of the Keldysh contour: it corresponds, apart 
from a prefactor,
to cutting the diagrams and putting the loose ends on the mass-shell. 
The resulting two off-diagonal processes conserve energy (tree structure
of the diagrams)
and can be read as 
spontaneous emission from atom $B$ after a resonant photon exchange (left
diagram) and as emission from atom B stimulated by a photon from atom A 
(right diagram).

To provide a more transparent physical interpretation,
we bring the above result into a form involving retarded and advanced
response functions.
These are related to the Keldysh propagators by 
$\alpha_{11} = \alpha_R + \alpha_{1 2}$ \todo{HH: Changed sign \Large$\checkmark$}
and $D_{2 1} = D_{1 1} - D_A$
(see Refs.\refcite{Danielewicz_1984,vanLeeuwen_2006}). Remembering that 
for positive frequencies and $T = 0$, we have 
$D_{1 1}( \omega ) = D_R( \omega )$ and
$\alpha^g_{1 1}( \omega ) = \alpha^g_{R}( \omega )$,
we get
\chdid{[CH: factor $i$ added, plus sign corrected in last line]}
\todo{HH: Swapped back signs of middle lines}
\begin{align}
\Sigma^{ge}_{1 1} 
	= i \int_0^\infty \frac{d \omega}{2 \pi }
	\alpha^{g B}_{R}(\omega)
	 \left\{ \right.
& D_{R}(\omega, \V r_B, \V r_A) \alpha^{e A}_{R}( \omega ) D_{R}(\omega, \V r_A, \V r_B) 
	\nonumber\\
& {} + D_{R}(\omega, \V r_B, \V r_A) \alpha^{e A}_{1 2}( \omega ) D_{R}(\omega, \V r_A, \V r_B)
&(*) 
	\nonumber\\
&{} - D_{R}(\omega, \V r_B, \V r_A) \alpha^{e A}_{1 2}( \omega ) D_{R}(\omega, \V r_A, \V r_B) 
&(*) 
	\nonumber\\
& {} + D_{R}(\omega, \V r_B, \V r_A) \alpha^{e A}_{1 2}( \omega ) 
D_{A}(\omega, \V r_A, \V r_B) 
\label{eqn:dressed_photon_3} 
\left. \right\}
~.
\end{align}
The two lines indicated by the asterisk $(*)$ cancel out exactly.
The first line gives an integrand regular in the upper right quadrant and
can be rotated onto the imaginary axis; it does not yield any oscillating
contributions. Within the two-level approximation, one has 
$\alpha_{R}^e = -\alpha_{R}^g$ [see Eq.\,(\ref{eqn:alpha11})], so that
the first term is equal, up to a global sign, to the ground-ground 
self-energy $\Sigma^{gg}_{1 1}$.
The last line gives a purely resonant contribution because
$\alpha^{e A}_{1 2}( \omega ) = 2\pi i \, \delta( \omega - \omega_A )$
according to \pr{eqn:pi_12}. It involves the modulus squared of the
(retarded) photon Green function because $D_A( \omega ) = D_R^*( \omega )$
at real frequencies. We thus rewrite Eqn.\eqref{eqn:dressed_photon_3} as
\todo{[CH: minus sign added in last term, consistent with imaginary
part; check real part.]}
\begin{align}
&\Sigma^{ge}_{1 1} 
	= - \Sigma^{gg}_{1 1}
  	-  
	\alpha^{g B}_{R}(\omega_A) \,
	|D_{R}(\omega_A, \V r_B, \V r_A)|^2 
	\label{eq:eg-potential-Keldysh}
\\
	&=
	{} - \hspace{-.2cm}\BoxGG{-2.5 ex}{.25\columnwidth} \hspace{-.25cm}
	- \hspace{-.2cm}\CrossGG{-2.5ex}{.25\columnwidth}
  	\hspace{-.25cm}
	- \left|\ResonanceInteraction{-2.5ex}{.2\columnwidth}\right|^2 
	\alpha^{g B}_{1 1}(\omega_A)
	~,
	\nonumber
\end{align}
where the last term can be identified as F\"orster resonant energy 
transfer\cite{Power_1983,Salam_2009,Forster_1949,Andrews_1989,Cohen_2003,Cohen_2003a} (FRET).
Rather than distinguishing absorption and emission, 
the above reordering has led to a separation of dispersion (nonresonant
or virtual photons) and FRET effects.
The resonant energy exchange makes the ground-state atom $B$ unstable,
and we recover the rate for FRET
\begin{equation}
\Gamma_{\rm FRET} = - 2 
\, {\rm Im} \bigl[ \Sigma^{g e}_{1 1} \bigr]
	=
	2 \pi \delta(\omega_B - \omega_A)
	|D_{R}(\omega_A, \V r_B, \V r_A)|^2 
	~,
\nonumber
\end{equation}
in full agreement with the Golden Rule. The
generalization of this rate to molecular emission and absorption spectra 
of finite width is straightforward and well-known \todo{good book to quote?}.
The FRET process comes along with a resonant two-body potential (the
real part of $\Sigma^{ge}_{1 1}$) that has been discussed in
Refs. \refcite{Power_1993}--\refcite{Sherkunov_2009}. This potential
does not oscillate spatially and lends itself to a simple semi-classical
picture: the polarization energy of atom $B$ in the time-averaged field
of a dipole source located at atom $A$, whose amplitude is fixed by the
transition dipole\cite{Power_1995}.
\chdid{this picture confirms the sign in Eq.\,(\ref{eq:eg-potential-Keldysh})
because the polarization energy is $- \alpha |E^2|$}

%

%
%
%
%
%
\subsection{Transient spatial oscillations of the resonant potential}
%
 %
It is instructive to evaluate the contribution that was cancelled from
Eq.\,(\ref{eqn:dressed_photon_3}) (upper line marked (*)). 
The explicit form of $\alpha^{eA}_{1 2}$ from \pr{eqn:pi_12} 
results in a spatially oscillating contribution to the self-energy
\chdid{[CH: only one sign]}\todo{HH: chose $-$}
\begin{align}
\Delta \Sigma^{eg}_{1 1} = - \alpha^{g B}(\omega_A) 
D_{1 1}(\omega_A, \V r_B, \V r_A)
D_{1 1}(\omega_A, \V r_A, \V r_B)
~.
	\label{eq:extra-term-with-oscillations}
\end{align}
Hence the first two (Feynman) diagrams in Eq.\,(\ref{eqn:dressed_photon_2})
give
\chdid{[CH: sign changed]}
%
\begin{align}
\BoxStandard{-2.5ex}{0.25\columnwidth} + \CrossStandard{-2.5ex}{0.25\columnwidth}
 = - \Sigma_{1 1}^{gg} + \Delta \Sigma_{1 1}^{eg}
 ~.
 \label{eqn:Feynman_result}
\end{align}
Since the potential $\Sigma_{1 1}^{gg}$ is monotonous, the extra 
term~(\ref{eq:extra-term-with-oscillations}) is responsible for
the spatially oscillating potential
found in Refs.\,\refcite{McLone_1965}--\refcite{Kweon_1993}
using equilibrium (Feynman) theory. 
If all four diagrams are taken into account, one reaches a fully equilibrated 
state and the interaction $\Sigma_{1 1}^{eg}$ 
{}[Eq.\,(\ref{eq:eg-potential-Keldysh})]
does no longer contain any spatial oscillations.
%
This is the result in Power and Thirunamachandran's 
calculations\cite{Power_1993,Power_1995}.
\todo{Did Sherkunov's 2005 paper give the oscillating result?}
\todo{What did the erratum of Kweon \& Lawandy from 1994 find? They
are more general because they take states above $|e\rangle$.} \todo{HH: To me it looks as if both groups (incl. the Erratum) found oscillations, can someone confirm?}

We suggest the following scenario to understand the physical relevance of
the two results, using the narrative of atom dressing\cite{Compagno_1995}.
The interchange of virtual (or non-resonant) photons responsible for 
the monotonous part of the potential is very fast. This short-time behaviour
is similar for both ground-state and excited atoms. The exchange of
near-resonant photons takes a longer time due to the small frequency
differences involved, and it leads to the equilibration of the initial state. 
This happens on the scale $1/\Gamma_{\rm FRET}$ set by the F\"orster rate.
This time appears as a lower limit for the validity of the equilibrated potential 
obtained from $\Sigma_{1 1}^{e g}$. An upper limit is set by the spontaneous
lifetime $1/\Gamma^e_0$ of the atom $A^*$
when photons appear in free space modes\cite{Sherkunov_2009,Cohen_2003}. The two time scales are well
separated at short distance (non-retarded range, typical for FRET 
experiments). 
%
%

The spatial oscillations of the potential~(\ref{eqn:Feynman_result})
may therefore appear as a transient effect right after the preparation. 
We speculate that they remain visible if the excited state is 
continuously replenished with a sufficiently high rate (larger than 
$\Gamma_{\rm FRET}$). 
This would require a time-dependent perturbative calculation similar to
Ref.\refcite{Power_1983} that is beyond
the scope of this paper.
We note that a spatially oscillating dispersion interaction has been found at
short times 
in the somewhat analogous situation of an excited molecule close to a 
surface\cite{Ellingsen_2009}.
%
%
\section{Discussion and Conclusions}
The apparently differing predictions for the resonant part of the two-body potential between a ground-state and an excited atom have been a source of confusion for quite some time. We have reviewed this system in a non-equilibrium description (Keldysh closed time-path formalism) and identified how contributions that oscillate spatially disappear in such a description.
The two results may correspond to different physical setups or to situations separated by a characteristic time scale, which we related to the rate of energy transfer (FRET): the oscillatory potential
(Refs.\,\refcite{McLone_1965}--\refcite{Kweon_1993}) holding on very short time scales, and eventually evolving into the non-oscillatory one of Refs.\,\refcite{Power_1993}--\refcite{Sherkunov_2009}.

In molecular physics, F\"orster broadening of atomic or molecular spectra is well known, but there seems to be less literature on the forces that come along with such resonant exchanges of energy\cite{Malnev_1970}.
Actually, the term \emph{F\"orster force} was coined as late as 2003 by Cohen and Mukamel\cite{Cohen_2003,Cohen_2003a}.
These forces may, however, play an important role in systems that stand in the focus of recent research, 
ranging from cold atoms\cite{Anderson_1998,Mourachko_1998} to quantum dots, NV centers, and even biomolecules such as proteins and DNA.
In the latter context, it has been proposed \todo{HH: I don't know by whom!}
that diffusion-limited reactions between an excited and a ground-state reactant are facilitated by a state-selective force. The force would modify the motion of the molecules relative to undirected Brownian diffusion which may help to understand unusually high reaction rates\cite{Berg_1985}.
In biological systems in particular, the excited state may be pumped by an energy source (photoabsorption, chemical). Then the energy flow through the system will determine to what extent the scenario involving equilibration after excitation actually applies.
Such questions certainly deserve further investigation, and underline that talking about unstable states as in dispersion interactions between atoms or molecules requires a careful characterization of the \emph{excited state} in question.

We thank R. Behunin, S. Buhmann, G. Cammarata, F. Intravaia, V. Mkrtchian, 
R. Passante,   G. Pieplow, J. Preto, S. Scheel, and S. Spagnolo for helpful discussions. 
Partial financial support by DFG, ERC, GIF, and DAAD is acknowledged.
Participation in QFExt11 was made possible by ERC and DAAD.
All diagrams were created using Jaxodraw\cite{Binosi_2004}.
\begin{appendix}
\section{Atom fields and propagators}
\label{sec:polarizations}
Atoms are considered distinguishable by an index $n = A, B$, with an internal two-level structure of energies $\epsilon_{g,e}^{n}$. 
The results will only depend on the Bohr frequencies
$\omega^n \equiv \epsilon_e^n - \epsilon_g^n$~.
%
The annihilation operator $\psi_a^n( x )$ evolves in the interaction picture
proportional to $\exp( -i \epsilon_{a}^{n} t )$ ($a = g, e$).
%
Atoms are considered immobile and pointlike so that the position
dependence is carried by the field operator 
in the interaction 
Hamiltonian~(\ref{eq:dipole-coupling-Hamiltonian}).
%
%
%

%
%
The bare (leading order) propagator for a ground-state atom is given by
\begin{align}
g_{1 1}^{n} ( x', x )
	&=
	-i \langle \psi_g^{n} (x') \psi_g^{n \dagger} (x) \rangle
	\theta(t' - t)
	&=
	\int \frac{d \omega}{2 \pi} \,
	\frac{e^{-i \omega (t' - t)}}
	{\omega - \epsilon_g^{n} + i\, 0^+}
\;.
\label{eq:g-propagator}
\end{align}
%
%
For an excited atom, $e_{1 1}^n$ is obtained using $\epsilon_e^n$ in place
of $\epsilon_g^n$.
%
The step function $\theta( t' - t )$ arises because in this non-relativistic
theory, there are no anti-particles.
In the frequency representation, we thus have the Feynman rules
\todo{harmonize diagrams}
\begin{align}
g_{1 1}(\omega) =  \frac{1}{\omega - \epsilon_g + i\, 0^+} = 
\includegraphics*[width=0.1\textwidth]{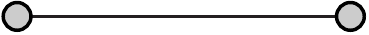}%
	\;, \quad 
e_{1 1}(\omega) = \frac{1}{\omega - \epsilon_e + i\, 0^+}  = 
\raisebox{0ex}{\includegraphics*[width=0.1\textwidth]{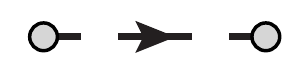}}
	\label{eq:g-and-e-propagator-11}
\end{align}
%

%
We also need Keldysh propagators for the atomic polarization operator
$P( x_n ) \equiv \psi^e(x_n) \psi^{g\dagger}(x_n) 
	+ \psi^g(x_n) \psi^{e\dagger}(x_n)$ that appears in the dipole 
interaction~(\ref{eq:dipole-coupling-Hamiltonian}).
The general correlation is 
\begin{equation}
	\alpha^{an}_{\alpha\beta}( x', x ) =
	i \langle T_c\{ P_\alpha( x'_n ) P_\beta( x_n ) \}\rangle_a
	\label{eq:def-polarization-correlation}
\end{equation}
for atom $n$ in state $a$. 
\todo{Please check for consistency.}
%
\begin{align}
\alpha_{1 1}^{e n} ( x', x )
	&=
	- \bigl(
	\langle e |
	\psi_e^{n\dagger} ( x' )
	\psi_e^n ( x )
	| e \rangle
	g^n_{1 1}( x', x )
	+
	x' \leftrightarrow x
	\bigr)
\\
	&=
	\int \frac{d \omega}{2 \pi} \,
	e^{-i \omega (t' - t)} \,
	\biggl(
	\frac {-1} {\omega^n + \omega + i\, 0^+}
	+
	\frac {-1} {\omega^n - \omega + i\, 0^+}
	\biggr)
\end{align}
The Feynman polarizability
$
\alpha_{1 1}^{e n}(\omega)
$
of the excited atom has a pole in the upper right quadrant of 
the $\omega$-plane which leads to a residue (``resonant contribution'') 
when integration
contours are shifted to the positive imaginary axis.
The non-equilibrium polarizability gives only a resonant contribution
\chdid{minus sign}
\begin{align}
\alpha_{1 2}^{e n}( x', x )
	&=
	i \langle e |
	\psi_e^{n\dagger} ( x )
	\psi_g^n( x )
	\psi_e^n ( x' )
	\psi_g^{n\dagger} ( x' )
	| e \rangle
	&=
	\int \frac {d \omega}{2 \pi}\,
	e^{-i \omega (t' - t)}\,
	2 \pi i \,
	\delta (\omega -  \omega^n)
\;.
\label{eqn:pi_12}
\end{align}
%
%
To illustrate the link to the retarded polarizability, we form the combination
\chdid{note the minus sign, conforming with the relation for $D$}
\begin{align}
\alpha^{en}_{1 1} ( x', x ) - \alpha^{en}_{1 2} ( x', x )
	&=
	-\int \frac {d \omega}{2 \pi}
	e^{-i \omega (t' - t)}
	\biggl(
	\frac{1}{\omega^n + \omega +i\, 0^+}
	+
	\frac{1}{\omega^n - \omega - i\, 0^+}
	\biggr)
\nonumber
%
%
%
\\&	=
	\int \frac {d \omega}{2 \pi}
	e^{-i \omega (t' - t)}
	\alpha^{e}_{\rm R} (\omega, \V{r}_1, \V{r}_2)
	\;,
\label{eqn:p11_def_b}
\end{align}
which has poles in the lower half-plane only, as it should.
%
%
%
For ground-state atoms, replace $\omega^n \leftrightarrow -\omega^n$
in these expressions, cf. \pr{eqn:alpha11} for $\alpha^{gn}_{1 1}$.
%

%
\section{Green functions of the free electromagnetic field}
\label{sec:free_photons}
%
We use the following Keldysh--Green functions for the photon field.
To illustrate the formalism, we allow in this appendix for a thermal state
with inverse temperature $\beta$ whence 
\chdid{$x',x$ notation; sign in $\bar n( \pm \omega )$ fixed to conform
with (anti)normal ordered results} \todo{HH: \Large$\checkmark$}
\begin{alignat*}{4}
D_{1 1} (x', x)
	&=
	i\langle
	T\bigl\{
	E( x' ) E( x )
	\bigr\}
	\rangle_{\beta}
&\, =& ~~~~~\int\frac{d\omega}{2 \pi} e^{-i \omega( t' - t )}
D_{1 1} (\omega, \V r', \V r )
\;,
\\
D_{1 2} (x', x)
	&=
	i\langle
	E( x ) E( x' )
	\rangle_{\beta}
&=& ~~~ i \int\frac{d\omega}{2 \pi} e^{-i \omega( t' - t )}
 2 \bar{n}( \omega )\,\mbox{Im}[D_{R} (\omega, \V r', \V r )]
\;,
\\
D_{2 1} (x', x)
	&=
	i\langle
	E( x' ) E( x )
	\rangle_{\beta}
&=& -i \int\frac{d\omega}{2 \pi} e^{-i \omega( t' - t )}
 2 \bar{n}(-\omega)\,\mbox{Im}[D_{R} (\omega, \V r', \V r )]
\;.
\end{alignat*}
Note that the thermal occupation number becomes
$n(\pm \omega) \to \mp \theta(\mp \omega)$ 
\chdid{factor $2$ removed} \todo{HH: \Large$\checkmark$}
at zero temperature. The Fourier transforms $D_{1 2}(\omega)$ 
and $D_{2 1}(\omega)$ are then only supported by 
negative / positive frequencies, respectively. In the rest of the paper
we always use this limit. The general case given above follows from
the fluctuation-dissipation theorem~\cite{vanLeeuwen_2006,Agarwal_1975a}.
%
%
The link to the retarded and advanced Green functions is provided by 
the relations %
$D_{\rm R}	= D_{1 1}  - D_{1 2}$ and  
$D_{\rm A}	= D_{1 1}  - D_{2 1}. 
$ 
%
%
The Feynman propagator in the frequency domain is, therefore,
\begin{align}
D_{1 1}(\omega, \V r', \V r ) 
 = 
 \SimplePhoton{-0.1ex}{0.07\textwidth}
%
= \mbox{Re} [D_R(\omega, \V r', \V r ] + i \coth\left(\beta \omega/2\right)\mbox{Im} [D_R(\omega, \V r', \V r )] ~.
	\label{eq:photon-11-finite-T}
\end{align}
The retarded Green tensor in a linear and isotropic medium 
depends only on the difference $\V r' - \V r$ and is given by
\begin{align}
D_{R}(\omega, \V r, \V 0 ) = & \frac{4 \pi}{3} \delta(r) \mathbbm{1} +  \frac{\omega^2}{c^2}\frac{ e^{i k r}}{r}    
\left[\left(1 + \frac{i}{k r} - \frac{1}{k^2 r^2}\right)\mathbbm{1} 
- \left(1 +  \frac{3i}{k r} - \frac{3}{k^2 r^2}\right) 
\hat{\V r} \otimes \hat{\V r}  \right]~,
\nonumber
\end{align}
where $\hat{ \V r } = \V r / |r|$, and 
$k = \sqrt{\varepsilon(\omega) \mu(\omega)} \omega / c$ 
(${\rm Im}[k] > 0$) is the wave vector in the medium.
Obviously, $[ D_R(\omega, \V r, \V 0) ]^2$ oscillates at half 
the medium wavelength, while $|D_R(\omega, \V r, \V 0)|^2$ does not. 
In the paper, $D_R( \omega, \V r_B, \V r_A )$ is evaluated by contracting
the tensor above from left and right with the transition dipoles
${\bf d}^B$ and ${\bf d}^A$.
\vspace{-.5ex}
%
%

%
\end{appendix}

{\footnotesize

\begin{thebibliography}{30}

\bibitem{Casimir_1948}
H.~B. Casimir and D. Polder, Phys. Rev. {\bf 73},  360  (1948).

\bibitem{Power_1983}
E.~A. Power and T. Thirunamachandran, Phys. Rev. A {\bf 28},  2671  (1983).

\bibitem{McLachlan_1963}
A. McLachlan, Proc. Roy. Soc. Series A. {\bf 271},  387  (1963).

\bibitem{Craig_1998}
D. Craig and T. Thirunamachandran, {\em {Molecular quantum electrodynamics}}
  (Dover Publications, Mineola, New York, 1998).

\bibitem{McLone_1965}
R.~R. McLone and E.~A. Power, Proc. Roy. Soc. A. {\bf 286},  573  (1965).

\bibitem{Philpott_1966}
M.~R. Philpott, Proc. Phys. Soc. {\bf 87},  619  (1966).

\bibitem{Kweon_1993}
G.-I. Kweon and N.~M. Lawandy, Phys. Rev. A {\bf 47},  4513  (1993), erratum:
  Phys. Rev. A {\bf 49}, 2205 (1994).

\bibitem{Power_1993}
E. Power and T. Thirunamachandran, Phys. Rev. A {\bf 47},  2539  (1993).

\bibitem{Power_1995}
E. Power and T. Thirunamachandran, Phys. Rev. A {\bf 51},  3660  (1995).

\bibitem{Sherkunov_2007}
Y. Sherkunov, Phys. Rev. A {\bf 75},  012705  (2007).

\bibitem{Sherkunov_2009}
Y. Sherkunov, Phys. Rev. A {\bf 79},  032101  (2009).

\bibitem{Salam_2009}
A. Salam, {\em {Molecular Quantum Electrodynamics}} (Wiley, Hoboken, New
  Jersey, 2009).

\bibitem{Danielewicz_1984}
P. Danielewicz, Annals of Physics {\bf 152},  239   (1984).

\bibitem{Sherkunov_2005}
Y. Sherkunov, Phys. Rev. A {\bf 72},  052703  (2005).

\bibitem{Schiefele_2010}
J. Schiefele and C. Henkel, Phys. Rev. A {\bf 82},  023605  (2010).

\bibitem{Schiefele_2011}
J. Schiefele, Ph.D. thesis, {Universit\"at Potsdam}, 2011.

\bibitem{Compagno_1995}
G. Compagno, R. Passante, and F. Persico, {\em {Atom-Field interactions and
  dressed atoms}} (Cambridge Univ. Press, Cambridge, 1995).

\bibitem{Wylie_1985}
J.~M. Wylie and J.~E. Sipe, Phys. Rev. A {\bf 32},  2030  (1985).

\bibitem{vanLeeuwen_2006}
R. van Leeuwen, N. Dahlen, G. Stefanucci, C.-O. Almbladh, and U. von Barth,  in
  {\em Time-Dependent Density Functional Theory}, Vol.~706 of {\em Lecture
  Notes in Physics}, edited by M. Marques, C. Ullrich, F. Nogueira, A. Rubio,
  K. Burke, and E. Gross (Springer, Berlin/Heidelberg, 2006), pp.\ 33--59.

\bibitem{Forster_1949}
T. F{\"o}rster, Z. Naturforsch. A {\bf 4},  321  (1949).

\bibitem{Andrews_1989}
D. Andrews, Chem. Phys. {\bf 135},  195  (1989).

\bibitem{Cohen_2003}
A.~E. Cohen and S. Mukamel, Phys. Rev. Lett. {\bf 91},  233202  (2003).

\bibitem{Cohen_2003a}
A.~E. Cohen and S. Mukamel, J. Phys. Chem. A {\bf 107},  3633  (2003).

\bibitem{Ellingsen_2009}
S.~A. Ellingsen, S.~Y. Buhmann, and S. Scheel, Phys. Rev. A {\bf 79},  052903
  (2009).

\bibitem{Malnev_1970}
V.~M. Mal'nev and S.~I. Pekar, Sov. Phys. JETP {\bf 31},  597  (1970).

\bibitem{Anderson_1998}
W.~R. Anderson, J.~R. Veale, and T.~F. Gallagher, Phys. Rev. Lett. {\bf 80},
  249  (1998).

\bibitem{Mourachko_1998}
I. Mourachko, D. Comparat, F. de~Tomasi, A. Fioretti, P. Nosbaum, V.~M. Akulin,
  and P. Pillet, Phys. Rev. Lett. {\bf 80},  253  (1998).

\bibitem{Berg_1985}
O. Berg and P. von Hippel, Ann. Rev. Biophys. Biophys. Chem. {\bf 14},  131
  (1985).

\bibitem{Binosi_2004}
D. Binosi, J. Collins, C. Kaufhold, and L. {Theu\ss l}, Comp. Phys. Commun.
  {\bf 180},  1709  (2009).

\bibitem{Agarwal_1975a}
G.~S. Agarwal, Phys. Rev. A {\bf 11},  230  (1975).

\end{thebibliography}

}
\end{document}